\documentclass[a4paper,twocolumn]{article}
\usepackage[dvips]{graphicx}
\begin{document}

\title{Characterization of entanglement using sum
uncertainty relations for N-level systems}

\author{Holger F. Hofmann and Shigeki Takeuchi\\
PRESTO, Japan Science and Technology Corporation (JST),\\  
Research Institute for Electronic Science,\\ 
Hokkaido University, Sapporo 060-0812\\
Tel/Fax: 011-706-2648\\ 
e-mail: h.hofmann@osa.org}

\date{}

\maketitle

\begin{abstract}
The efficient experimental verification of entanglement
requires an identification of the essential physical
properties that distinguish entangled states from non-entangled
states. Since the most characteristic feature of entanglement
is the extreme precision of correlations between spatially 
separated systems, we propose a quantitative
criterion based on local uncertainty relations (quant-ph/0212090). 
Some basic sum uncertainty relations for N-level systems
are introduced and the amount of entanglement that can be 
verified by violations of the corresponding local uncertainty 
limit is discussed.
\\[0.2cm]
\end{abstract}

\section{Introduction}
As the word itself suggests, entanglement is one of the 
most mysterious aspects of quantum physics. Although it
is by now possible to verify the predictions of 
entanglement theory in a variety of experiments, there
remains a considerable gap between the formal definition
of entanglement and the observable effects that are 
associated with this property. The formal definition
of entanglement is that any quantum state of two systems 
that cannot be decomposed into a mixture of product states 
of the two systems is an entangled state. However, the 
decomposition of a mixed state does not represent any real 
physical process, it is only one of many possible 
interpretations of the origin of the statistical mixture. 
Consequently, the physical meaning of this definition
is unclear. Indeed, it remains a highly non-trivial task
to identify physical properties that can be used to 
distinguish mixed entangled states from separable ones
\cite{witness}. 

In order to identify some characteristic physical properties of
entanglement, it may be useful to reconsider the original
discussions of entanglement \cite{classics}. These discussions
focussed on the fact that maximally entangled states have
perfect correlations between every single observable property,
that is, the outcome of any measurement of a property A in
system one can be predicted by measuring a corresponding property
B in system two. In a classical system, this would not be 
surprising at all. System one is simply an exact copy of system
two. However, such a high level of precision appears to 
contradict the uncertainty principle. Indeed, complete
knowledge of the local quantum states of two systems requires 
that there are no correlations between the two systems, 
so that the uncertainties of the two pure states represent 
statistically independent fluctuations of the respective 
measurement outcomes.
Therefore, the correlations that exist in mixtures of local 
states can never overcome the limits set by local uncertainties.
Only the strong correlations between measurement fluctuations
associated with entangled states can
violate such local uncertainty limits. The definition
of local uncertainty relations can thus provide a directly
observable measure of entanglement \cite{Hof02}. This measure 
of entanglement naturally provides an evaluation of the 
usefulness of an entangled state for applications such as 
quantum teleportation and dense coding,
where the main purpose of entanglement is the suppression of 
uncertainty related noise in the transfer of information 
\cite{Hof02b}. It may indeed
be an interesting question whether the useful properties of
entanglement really arise from the theoretical lack of 
separability, or if this formal requirement of entanglement
is only a necessary condition for the improvement of precision
described by the violation of local uncertainties.  

There remains one fundamental problem that has to be solved
in order to derive measures of entanglement from local
uncertainty relations. Previously, uncertainty arguments were
mostly restricted to continuous variables, where a 
non-vanishing uncertainty limit for the product of position
and momentum uncertainties provides a useful definition of
local uncertainties. Unfortunately, such product uncertainties
do not provide any real uncertainty limits for N-level systems
because the product uncertainties will always be zero for 
eigenstates of one of the two properties. It is therefore
necessary to find a different formulation of uncertainty
limits for the properties of N-level systems. As we have 
pointed out recently, such uncertainty limits can be 
obtained by using the sum of uncertainties instead of 
their product \cite{Hof02}. It is then possible to obtain 
sum uncertainty relations for any set of operators,
although in general, the determination of the correct 
uncertainty limit may be quite difficult.
In the following, we present some typical examples of sum 
uncertainty relations and discuss their applications to 
the experimental characterization and quantification of 
entanglement.

\section{Uncertainty relations for N-level systems}

The uncertainty $\delta A_i^2$ of a hermitian operator
property $\hat{A}_i$ is defined as the variance of the
measurement statistics of a precise measurement of 
$\hat{A}_i$. For a given density matrix $\hat{\rho}$,
this uncertainty is given by
\begin{equation}
\delta A_i^2 = \mbox{Tr}\{\rho \hat{A}_i^2\}
- \left(\mbox{Tr}\{\rho \hat{A}_i\}\right)^2.
\end{equation}
The uncertainty of $\hat{A}_i$ can only be zero if 
$\rho$ represents an eigenstate of $\hat{A}_i$.
Therefore, the sum of all uncertainties for a given
set of operators $\{\hat{A}_i\}$ can only be zero 
if all the operators in the set have at least one
common eigenstate. In all other cases, there exists
a non-trivial uncertainty limit given by
\begin{equation}
\label{eq:basic}
\sum_i \delta A_i^2 \geq U.
\end{equation}

In principle, there are infinitely many sum uncertainty
relations, since any set of operators may be used.
However, in most cases, the physical situation considered
will determine which choices are useful. 
In the following, we will focus on the angular momentum 
operators $\hat{L}_x$, $\hat{L}_y$, and $\hat{L}_z$ of
a spin $l$ systems. In general, this spin algebra can 
be defined for any N-level system, where $N=2 l + 1$. 
In many cases, the components of the spin algebra will
even have a clear physical meaning. For example, 
these operators can also be used to describe the Stokes 
parameters $\hat{S}_i$ of an n photon state, where $n=2 l$ 
and $\hat{S}_1= 2 \hat{L}_x$, $\hat{S}_2= 2 \hat{L}_y$, 
and $\hat{S}_3= 2 \hat{L}_z$.
The uncertainty relation for all three spin components
of an N-level system can be derived directly from
the properties of the operators, 
\begin{eqnarray}
\hat{L}_x^2 + \hat{L}_y^2 + \hat{L}_z^2 &=& l (l+1)
\nonumber \\
\mbox{and} \hspace{1cm} \langle \hat{L}_i \rangle^2 &\leq& l^2.
\end{eqnarray}
Since the spin system is isotropic, the expectation value limit 
of one component is enough to define the limit for the 
total averaged spin vector. The uncertainty relation then reads
\begin{equation}
\label{eq:SpinUR}
\delta L_x^2 +\delta L_y^2 +\delta L_z^2 \geq l.
\end{equation}
Likewise, the uncertainty relation for the Stokes parameters
of an $n$ photon system reads
\begin{equation}
\label{eq:StokesUR}
\delta S_1^2 +\delta S_2^2 +\delta S_3^2 \geq 2 n.
\end{equation}
In the case of a single photon ($n=1$), the Stokes parameters 
are equivalent to the Pauli matrices. This means that the single 
photon polarization state can be interpreted more generally 
as a physical implementation of a single qubit. 
Relation (\ref{eq:StokesUR}) is therefore particularly
useful for the description of quantum bits. In particular,
the uncertainty limit of two at $n=1$ expresses the fact that 
information can only be encoded in a single component of the 
spin vector.

It is also of interest to know the uncertainty relations
for only two spin components. However, it is a non-trivial
task to derive the proper uncertainty limit. Here, we would
only like to note that the best strategy to minimize the
uncertainty of $\hat{L}_x$ and $\hat{L}_y$ seems to be 
the application of spin squeezing to a coherent spin state,
e.g. the eigenstate of $\hat{L}_x$ with $L_x=+l$.
Spin squeezing can then redistribute the quantum fluctuations
of the state from $\hat{L}_y$ to $\hat{L}_z$ until the gradual
increase in $\delta L_x^2$ prevents any further reduction in
the sum uncertainty. However, care must be taken to ensure 
that the minimum obtained is actually the global minimum for 
all possible quantum states.

For two level systems, it is still very easy to obtain the
uncertainty limit for $\hat{L}_x$ and $\hat{L}_y$, since
every eigenstate of a spin component has the same maximal
amount of uncertainty in the other two components and a
redistribution of uncertainty by spin squeezing is not 
possible. The uncertainty relations therefore read
\begin{eqnarray}
\label{eq:N=2}
N=2 \hspace{0.3cm}\mbox{uncertainties}:&&
\nonumber \\[0.2cm]
\delta L_x^2 +\delta L_y^2 &\geq& \frac{1}{4}
\\
\label{eq:qubit}
\mbox{or}\hspace{0.5cm}
\delta S_1^2 +\delta S_2^2 &\geq& 1.
\end{eqnarray}
As mentioned above, the relation for the Stokes parameters 
also represents the uncertainty for two of the three Pauli
matrices of a general quantum bit. Therefore, relation 
(\ref{eq:qubit}) is a very compact formulation of the 
limitations of encoding and readout for any quantum 
communication scheme that encodes information in the 
eigenstates of both $\hat{S}_1$ and $\hat{S}_2$.

For three level systems, the uncertainty distribution is 
variable, and the minimum uncertainty state must be determined
by optimizing this distribution. However, the three dimensional
Hilbert space is still small enough to allow an analytical
determination of the global maximum. The result we have obtained
was first reported in \cite{Hof02} and reads
\begin{eqnarray}
\label{eq:N=3}
N=3 \hspace{0.3cm}\mbox{uncertainties}:&&
\nonumber \\[0.2cm]
\delta L_x^2 +\delta L_y^2 &\geq& \frac{7}{16}
\\[0.2cm]
\label{eq:qubit}
\mbox{or}\hspace{0.5cm}
\delta S_1^2 +\delta S_2^2 &\geq& \frac{7}{4}.
\end{eqnarray}
As suggested above, the minimum uncertainty states for this
relation are spin squeezed states with average spins in the
xy-plane. In the $\hat{L}_z$ basis, these minimal uncertainty
states read 
\begin{equation}
\mid \phi \rangle =
\frac{\sqrt{5}}{4} e^{-i \phi} \mid -1 \rangle
+  \frac{\sqrt{6}}{4} \mid 0 \rangle + 
\frac{\sqrt{5}}{4} e^{+i \phi} \mid +1 \rangle.
\end{equation}
It may be interesting to note that these states are quite
obviously not minimal uncertainty states of relation
(\ref{eq:SpinUR}). The notion of minimal uncertainty states
is therefore strongly dependent on the uncertainty relation
selected for their definition. In general, all pure states
are minimal uncertainty states for some set of uncertainty
relations. 

\section{Formulation of local uncertainty limits}
\label{sec:LUR}

It is now possible to construct local uncertainty relations
from any selection of basic uncertainties. As mentioned in
the introduction, the uncertainties of local states cannot
be correlated. Therefore, the uncertainty sums of any set
of joint properties $\{\hat{A}_i + \hat{B}_i\}$ cannot be lower
than the sum of the local uncertainty limits for
$\{\hat{A}_i\}$ and $\{\hat{B}_i\}$,
\begin{eqnarray} 
\mbox{limit for system A:}&&\sum_i \delta A_i^2 \geq U_A
\nonumber \\
\mbox{limit for system B:}&&\sum_i \delta B_i^2 \geq U_B
\nonumber \\
\mbox{local limit for A+B:} &&
\nonumber \\[0.2cm]
\sum_i \delta (A_i+B_i)^2 &\geq& U_A+U_B
\end{eqnarray}
Any violation of this local uncertainty limit indicates that
systems A and system B are entangled. (For a simple
mathematical proof of this property, see \cite{Hof02}.)

In general, it
is not even necessary to choose the same type of uncertainty
relation in each system. For example, it is possible to 
determine local uncertainty limits for $N \times M$ systems
simply by selecting one N-level uncertainty and one M-level
uncertainty. However, the basic principle of local uncertainty
violations is best illustrated by symmetric relations,
where the operators $\hat{A}_i$ and $\hat{B}_i$
describe the same physical properties in system A and system B,
respectively. For general $N \times N$ entanglement, such
a local uncertainty limit can be obtained from relation
(\ref{eq:SpinUR}) or (\ref{eq:StokesUR}), respectively. 
According to these uncertainty limits, separable states
must fulfill the conditions
\begin{eqnarray}
\sum_{i=x,y,z} \delta(L_i(A)+L_i(B))^2
\geq 2 l,
\\
\sum_{i=1}^{3}
\delta(S_i(A)+S_i(B))^2
\geq 4 n. 
\end{eqnarray}
The maximal violation of these local uncertainty relations
is obtained for the singlet state, which has a total
uncertainty of zero. Consequently, these uncertainty relations
provide a measure of how close any given state 
is to the maximal entanglement described by the singlet
state of the two spin $l$ systems. This closeness can be
quantified using the relative violation of local uncertainty
\cite{Hof02},
\begin{eqnarray}
\label{eq:CL3}
C_{L3} &=& 1-\frac{\sum_{i=x,y,z} \delta(L_i(A)+L_i(B))^2}{2 l}
\\
\label{eq:CS3}
C_{S3} &=& 1-\frac{\sum_{i=1}^{3}
\delta(S_i(A)+S_i(B))^2}{4 n}.
\end{eqnarray}
These measures of entanglement allow a direct experimental 
evaluation of the errors in entanglement generation, especially
if the experiment is intended to generate pure singlet state
entanglement. No special assumptions about any detailed 
properties of the actual density matrix are necessary, 
although it is of course interesting to study the effects
of some typical errors such as decoherence effects on the 
relative violation of local uncertainties. For example, the
addition of white noise represents a kind of worst case
scenario, 
\begin{equation}
\label{eq:rhown}
\hat{\rho}(p_W) = 
(1\!-\!p_W) \mid\! \mbox{sing.}\rangle \langle \mbox{sing.}\!\mid
+  p_W \frac{\hat{1}\otimes\hat{1}}{N^2} 
\end{equation}
The relation between the noise level $p_W$ and
the relative violation of local uncertainty then reads
\begin{equation} 
\label{eq:Cwn}
C_{L3/S3} = \! 1 - p_W \frac{N+1}{2}.
\end{equation}
It may be interesting to note that the effect of white 
noise errors on the relative violation of local uncertainty
increases with the number of level $N$. However, it should
be remembered that white noise errors become less and less
likely in larger systems, as the nearly macroscopic properties
of many level systems become easier to control. This example
therefore also illustrates the problems associated with the
choice of a noise model in larger Hilbert spaces.

\section{Application to $2\times 2$ entanglement}

The most simple case of entanglement is that between
a pair of two level systems. Since this kind of 
entanglement has been studied extensively in the 
context of application in quantum information
technologies, much is known about the fundamental
properties of such entanglement. Nevertheless, 
the application of local uncertainty relations can
be very useful in the study of $ 2\times 2 $ entanglement, 
since it provides a greatly simplified access to some
of the most characteristic features of this kind of
entanglement.

In particular, we can consider the case of a pair of 
entangled photons with no local polarizations and
different degrees of correlation between identical
components of the Stokes vector.
Such a state can be expressed as a mixture of 
four Bell states,
\begin{eqnarray}
\label{eq:2state}
\hat{\rho}(p_S;p_i) &=& 
p_S \mid S \rangle \langle S \mid +
p_1 \mid T1 \rangle \langle T1 \mid 
\nonumber \\ &+& \!
p_2 \mid T2 \rangle \langle T2 \mid +
p_3 \mid T3 \rangle \langle T3 \mid,
\end{eqnarray}
where $\mid S \rangle$ is the singlet state and the 
three states $\mid Ti \rangle$ are the triplet states
defined by 
\begin{equation}
(\hat{S}_i(A)+\hat{S}_i(B))\mid Ti \rangle = 0.
\end{equation}
The quantum state $\hat{\rho}(p_S;p_{i})$ is entangled if
the probability of one of the maximally entangled
states exceeds one half. Specifically, if $p_S>1/2$,
the concurrence of the quantum state is given by 
\begin{equation}
C = 2 p_S - 1.
\end{equation}
The concurrence is a precise measure of the total 
entanglement in the $2\times 2$ system and is directly 
related to the entanglement of formation.
It is therefore interesting to compare this basis
independent measure of entanglement with the relative
violation of local uncertainties.

The properties of the quantum state $\hat{\rho}(p_S;p_i)$ 
can be determined completely by measuring the three 
uncertainties $\delta(S_i(A)+S_i(B))^2$.
For the quantum state $\hat{\rho}(p_S;p_i)$,
these uncertainties are given by
\begin{equation}
\delta(S_i(A)+S_i(B))^2 = 4-4(p_S+p_i).
\end{equation}
The corresponding violation of local uncertainty 
given by (\ref{eq:StokesUR}) reads
\[
\sum_{i=1}^3 \delta(S_i(A)+S_i(B))^2 = 8 (1-p_s) < 4,
\]
\begin{equation}
C_{S3} = 2 p_S - 1 = C. 
\end{equation}
In this case, the relative violation of local uncertainties
is indeed equal to the concurrence. The local
uncertainty relation (\ref{eq:StokesUR}) therefore  
identifies the physical properties that "concur". 
Local uncertainty relations can thus specify 
the actual physical situation represented by a given 
entangled state.

In order to reduce the experimental effort, it may also
be desirable to apply the local uncertainty limit for
two components,
\begin{equation}
\delta(S_1(A)+S_1(B))^2 + \delta(S_2(A)+S_2(B))^2 \geq 2.
\end{equation}
In this case, there is a slight difference between the
relative violation of local uncertainty and the concurrence,
\[
\sum_{i=1}^2 \delta(S_i(A)+S_i(B))^2 = 8 (1-p_s+p_3) < 2,
\]
\begin{equation}
C_{S2} = 2 p_S - 1 - 2 p_3 \leq C. 
\end{equation}
The relative violation $C_{S2}$ of the local two component 
uncertainty may therefore serve as a lowest limit of the
actual concurrence. This lowest estimate corresponds to the 
"worst case scenario" that $p_3$ is zero and that the
uncertainty in $\hat{S}_3(A)+\hat{S}_3(B)$ has the maximal
value consistent with the other two uncertainties.
It is thus possible to identify the quantum state
associated with this lowest estimate from only two 
measurement settings.

In the context of photon pair entanglement, it may also be
useful to identify the precise relationship between 
coincidence counts and the joint uncertainties
$\delta(S_i(A)+S_i(B))^2$. Experimentally, the correlation
between Stokes parameter components is usually evaluated
by comparing the maximal rate of coincidence counts 
observed at anti-correlated polarizer settings with the
minimal rate of coincidence counts at correlated settings.
The normalized difference between the two defines the 
visibility $V_i$ for the polarization component $i$.
In the absence of local polarizations, this measurement
result is related to the uncertainty by
\begin{equation}
\delta(S_i(A)+S_i(B))^2 = 2(1-V_i).
\end{equation}
It is therefore possible to derive an estimate of the
concurrence directly from a measurement of the 
visibilities $V_1$ and $V_2$,
\begin{equation}
C \geq V_1+V_2-1.
\end{equation}
The visibilities thus allow an even greater simplification
of the experimental evaluation of $2 \times 2$ entanglement.
However, it should be pointed out that local uncertainties
generally do include the effect of non-vanishing local 
polarizations and are therefore a more precise measure of 
the actual entanglement correlations. 

\section{Application to $3\times 3$ entanglement}

Recently, the experimental generation of entanglement 
in parametric downconversion has been extended to 
pairs of three level systems \cite{3x3}. However, the
great variety of possible error sources in such systems
has made a precise analysis of the experimental results
very difficult. We therefore believe that the local 
uncertainty relations can provide a powerful tool for
a quantitative estimate of the amount of entanglement 
created in such experiments. 

As mentioned in section \ref{sec:LUR}, one possible
type of error in the creation of $3 \times 3$ entanglement
could be white noise of the form given by $\hat{\rho}(p_W)$
in equation (\ref{eq:rhown}). For three level systems, the
corresponding relative violation of the three component
local uncertainty relation (\ref{eq:SpinUR}) reads
\[
\sum_{i=x,y,z} \delta(L_i(A)+L_i(B))^2 = 4 p_W < 2,
\]
\begin{equation}
C_{L3} = 1 - 2 p_W. 
\end{equation}
It is also possible to apply the two component uncertainty
using the uncertainty limit given by relation (\ref{eq:N=3}),
\[
\sum_{i=x,y} \delta(L_i(A)+L_i(B))^2 = \frac{8}{3} 
p_W < \frac{7}{8},
\]
\begin{equation}
C_{L2} = 1 - \frac{64}{21} p_W. 
\end{equation}
White noise therefore reduces the relative violation of 
the two component local uncertainty about 1.5 times 
faster than it reduces the relative violation of the 
three component uncertainty. However, this sensitivity
ratio between the three component and the two component
uncertainties can be quite different for less symmetric 
noise sources.

An alternative source of error could be decoherence in the
$\hat{L}_x$ basis, given by
\begin{eqnarray} 
\lefteqn{\hat{\rho}(p_D) =}
\nonumber \\ && 
 (1-p_D)
\mid\! \mbox{sing.}\rangle \langle \mbox{sing.}\!\mid +
\frac{p_D}{3}\big(
\mid\! -1;+1 \rangle \langle -1;+1 \!\mid 
\nonumber \\ && +
\mid\! 0;0 \rangle \langle 0;0 \!\mid +
\mid\! +1;-1 \rangle \langle +1;-1 \!\mid
\big).
\end{eqnarray}
In this case, the uncertainty in 
$\hat{L}_x(A)+\hat{L}_x(B)$ always remains zero, while 
the other two uncertainties increase as they did for 
white noise.
The three component local uncertainty then reads
\[
\sum_{i=x,y,z} \delta(L_i(A)+L_i(B))^2 = \frac{8}{3} p_D < 2,
\]
\begin{equation}
C_{L3} = 1 - \frac{4}{3} p_D. 
\end{equation}
As expected, the reduction of the local uncertainty violation 
by decoherence is only $2/3$ of that caused by the same 
amount of white noise.
Likewise, the effect of decoherence on the two component
local uncertainty violation is reduced to one half,
\[
\sum_{i=x,y} \delta(L_i(A)+L_i(B))^2 = \frac{4}{3} p_D 
< \frac{7}{8},
\]
\begin{equation}
C_{L2} = 1 - \frac{32}{21} p_D. 
\end{equation}
As a result, the two component local uncertainty is nearly as
good at quantifying the remaining entanglement in the 
presence of decoherence as the corresponding three component
uncertainty. If decoherence is assumed to be the main
source of error, it is therefore sufficient to characterize
experimentally generated three level entanglement using
only two measurement settings.

\section{A short note on the interpretation of 
quantum statistics}

As shown in this paper, entanglement is a very
useful property because it represents strong non-local 
correlations that can overcome local uncertainty 
limits. However, the non-locality of these correlations
appears to be no different from the non-locality
of classical correlations. Since there is no
direct evidence for action at a distance, it may
be more prudent to interpret the collapse of the 
wavefunction as the reduction of a probability 
distribution based on new information. 
More mysterious effects such as the violation
of Bell's inequalities could then be interpreted
in terms of local non-classical correlations
\cite{negprop}.

The decomposition of density matrices into mixtures
of pure states unfortunately tends to suggest that
there might be a fundamental difference between quantum 
noise and classical noise, and that the collapse of
the wavefunction should be somehow different from the
selection of a classical subensemble. However, the 
uncertainty relations indicates that it may be more 
realistic to define only a quantitative limit - in the 
same way that the speed of light defines the relativistic 
limit, without suggesting a separation of velocities into 
a relativistic and a non-relativistic component. 
As mentioned in the introduction, the ambiguity in the 
decomposition of density matrices into pure state mixtures 
indicates that there is no physical meaning in an arbitrary
decomposition. The uncertainty principle can explain this 
observation by suggesting that even pure states are very noisy. 
For all practical purposes, pure states also represent 
probability distributions over many possible measurement 
outcomes, and there is no observable difference between classical 
noise and quantum noise. A separation of mixed states into
pure states therefore divides the total ensemble into a 
rather arbitrary choice of subensembles. However, a pure
state is not sufficiently "pure" to define the physical
properties of an individual representative of the total 
probability distribution. The interpretational problems of 
quantum mechanics arise from the lack of any fundamental set 
of noise free subensembles that could identify the physical 
properties of individual systems as measurement independent
"elements of reality".

\section{Conclusions}

The definition of uncertainty relations for arbitrary
sets of operators describing the physical properties
of N-level systems allows a generalized interpretation
of entanglement as a violation of local uncertainty
relations. The quantitative measure given by the
relative violation of local uncertainties seems to 
correspond well with more theoretical measures such
as the concurrence. Since the local uncertainties 
are defined as variances of actual measurement results
they greatly simplify the theoretical background needed
to device tests of experimental realizations of
entanglement sources. Moreover, the violation of local
uncertainties directly identifies the improvement of 
precision achieved by entanglement in applications such
as dense coding or quantum teleportation.


\end{document}